\documentclass[12pt,a4paper]{article}
\usepackage[T1]{fontenc}
\usepackage[latin1]{inputenc}
\usepackage{color}

\usepackage{booktabs}

\usepackage{graphicx}

\usepackage{amsfonts}
\usepackage{amsmath}
\usepackage{amssymb,stmaryrd}
\usepackage{theorem}
\usepackage{mathtools}
\usepackage{mathtools}
\usepackage{etoolbox}

\DeclareFontFamily{U}{matha}{\hyphenchar\font45}
\DeclareFontShape{U}{matha}{m}{n}{
<-6> matha5 <6-7> matha6 <7-8> matha7
<8-9> matha8 <9-10> matha9
<10-12> matha10 <12-> matha12
}{}
\DeclareSymbolFont{matha}{U}{matha}{m}{n}

\DeclareFontFamily{U}{mathx}{\hyphenchar\font45}
\DeclareFontShape{U}{mathx}{m}{n}{
<-6> mathx5 <6-7> mathx6 <7-8> mathx7
<8-9> mathx8 <9-10> mathx9
<10-12> mathx10 <12-> mathx12
}{}
\DeclareSymbolFont{mathx}{U}{mathx}{m}{n}

\DeclareMathDelimiter{\vvvert} {0}{matha}{"7E}{mathx}{"17}%

\theoremstyle{change}
\theorembodyfont{\slshape}
\newtheorem{satz}{Theorem}[subsection]
\newtheorem{thm}[satz]{Theorem}

\theorembodyfont{\upshape\small}
\newtheorem{bsp}[satz]{Example}
\newtheorem{bem}[satz]{Remark}

\theorembodyfont{\ttfamily}

\numberwithin{equation}{section}
\newcommand{\ba}{\begin{equation}}
\newcommand{\ea}{\end{equation}}

\newcommand{\fB}{\mbox{\boldmath $B$}}

\newcommand{\fx}{\mbox{\boldmath $x$}}
\newcommand{\fY}{\mbox{\boldmath $Y$}}
\newcommand{\fy}{\mbox{\boldmath $y$}}
\newcommand{\fyi}{\mbox{\scriptsize\boldmath $y$}}

\newcommand{\mA}{\mbox{\textup{\textbf{A}}}}

\newcommand{\mM}{\mbox{\textup{\textbf{M}}}}

\newcommand{\feta}{\mbox{\boldmath $\eta$}}
\newcommand{\fxi}{\mbox{\boldmath $\xi$}}
\newcommand{\ftheta}{\mbox{\boldmath $\theta$}}

\newcommand{\bin}{\textup{Bin}}

\newcommand{\pgf}{\textup{pgf}}

\newcommand{\bbn}{\mathbb{N}}
\newcommand{\bbr}{\mathbb{R}}
\newcommand{\bbz}{\mathbb{Z}}
\newcommand{\e}{\mathbb{E}}
\newcommand{\V}{\textrm{Var}}
\newcommand{\Cov}{\textrm{Cov}}

\newcommand{\iid}{i.\,i.\,d.}
\newcommand{\ie}{i.\,e., }
\newcommand{\eg}{e.\,g., }

\usepackage{dsfont}
\newcommand{\indfkt}{\mathds{1}}

\usepackage{natbib}

\usepackage{hyperref}

\allowdisplaybreaks

\usepackage{indentfirst}
\begin{document}



\parindent 0cm

\title{Mean-preserving Rounding Integer-valued ARMA Models}
\author{
Christian H.\ Wei\ss{}\thanks{
Helmut Schmidt University, Department of Mathematics and Statistics, Hamburg, Germany. E-Mail: \href{mailto:weissc@hsu-hh.de}{\nolinkurl{weissc@hsu-hh.de}}. ORCID: \href{https://orcid.org/0000-0001-8739-6631}{\nolinkurl{0000-0001-8739-6631}}.}
\and
Fukang Zhu\thanks{
School of Mathematics, Jilin University, 2699 Qianjin, Changchun 130012, China. E-Mail: \href{mailto:zfk8010@163.com}{\nolinkurl{zfk8010@163.com}}. ORCID: \href{https://orcid.org/0000-0002-8808-8179}{\nolinkurl{0000-0002-8808-8179}}.}\ \thanks{Corresponding author.}
}

\maketitle

\begin{abstract}
\noindent
In the past four decades, research on count time series has made significant progress, but research on $\bbz$-valued time series is relatively rare. Existing $\bbz$-valued models are mainly of autoregressive structure, where the use of the rounding operator is very natural. Because of the discontinuity of the rounding operator, the formulation of the corresponding model identifiability conditions and the computation of parameter estimators need special attention. It is also difficult to derive closed-form formulae for crucial stochastic properties. We rediscover a stochastic rounding operator, referred to as mean-preserving rounding, which overcomes the above drawbacks. Then, a novel class of $\bbz$-valued ARMA models based on the new operator is proposed, and the existence of stationary solutions of the models is established. Stochastic properties including closed-form formulae for (conditional) moments, autocorrelation function, and conditional distributions are obtained. The advantages of our novel model class compared to existing ones are demonstrated. In particular, our model construction avoids identifiability issues such that maximum likelihood estimation is possible. A simulation study is provided, and the appealing performance of the new models is shown by several real-world data sets.

\medskip


\end{abstract}

\section{Introduction}
\label{Introduction}
During the last decades, there was a lot of research on discrete-valued time series, see \citet{weiss18} for a survey, where the large majority of contributions focused on count time series, \ie quantitative time series having a range contained in the set of non-negative integers, $\bbn_0=\{0,1,\ldots\}$. Only during the last fifteen years \citep[see][]{li24}, there was increasing interest also on time series that have the full set of integers, $\bbz=\{\ldots, -1,0,1,\ldots\}$, as their range, \ie where also negative integer outcomes are possible. Since the (actually applicable) term ``integer-valued time series'' is often used in the context of count time series, it became practice to use ``$\bbz$-valued time series'' instead in such a case.
\medskip

The first model for $\bbz$-valued time series appears to be the autoregressive (AR) model by \citet{kim08}, where the traditional AR-recursion is adapted to the integer case by using so-called ``thinning operators'' \citep[see][]{scotto15}. Since that time, several thinning-based ARMA-like models (autoregressive moving average) have been proposed, commonly referred to as integer-valued ARMA (INARMA) models, see \citet{scotto15,li24} for surveys. These models, however, have the disadvantage that the computation of probabilities and likelihoods can get demanding as convolutions of distributions are involved. Another approach for $\bbz$-valued time series are rounded ARMA models as suggested by \citet{kachour09,kachour14}, \ie the outcome of the ARMA recursion is rounded to the next integer value (also see the related approach by \citet{kowal20}). In this case, however, it is difficult to derive closed-form formulae for crucial stochastic properties, in particular, moment formulae are not available. A third popular approach (besides several further proposals as surveyed by \citet{li24}) are $\bbz$-valued INGARCH models (integer-valued generalized autoregressive conditional heteroscedasticity), see \cite{xu22} and \citet{liu23}, but these only specify the one-step ahead conditional distribution while other stochastic properties are difficult to derive (if possible at all). So none of the proposed model classes is entirely satisfactory from a practical point of view.
\medskip

Therefore, in this work, we propose a new class of ARMA-like models for $\bbz$-valued time series, which combines ideas from both INARMA models and rounded ARMA models, and which is constructed in such a way that the main disadvantages of INARMA and rounded ARMA models are simultaneously avoided. Inspired by \citet{liu13}, the crucial idea is to use a particular type of randomized rounding operator (\ie which combines the randomness of a thinning operator with the rounding operation) together with the ARMA recursion. This randomized rounding operator is constructed such that we get simple closed-form formulae for (conditional) moments, autocorrelation function (acf), and conditional distributions.
\medskip

Definition and properties of this operator are presented in Section~\ref{A Mean-preserving Rounding Operator}, while the proposed time series models are discussed in Section~\ref{A Novel Class of Integer-valued ARMA Models}. In particular, the advantages of our novel model class compared to existing ones are demonstrated. Among others, our model construction avoids identifiability issues such that maximum likelihood estimation is possible; the corresponding finite-sample performance is investigated by simulations, see Section~\ref{Performance of Parameter Estimation} as well as the Supplement. In Section~\ref{Real-world Data Applications}, the practical benefit of our model class is demonstrated by three real-world data applications, namely to $\bbz$-valued time series from finance, demography, and ecology (three further data examples are provided by the Supplement). Finally, we conclude in Section~\ref{Conclusions} and outline directions for future research.

\section{A Mean-preserving Rounding Operator}
\label{A Mean-preserving Rounding Operator}
In this section, we define and investigate the randomized rounding operator, mapping real numbers from~$\bbr$ onto integers from~$\bbz$, which is later used for defining integer-valued ARMA-type time series models.
More precisely, we define $\langle\cdot\rangle: \bbr\to\bbz$ as a random operator which does a kind of mean-preserving rounding operation:
\ba
\label{rounding_operator}
\langle z\rangle\ :=\ \left\{\begin{array}{ll}
\lfloor z\rfloor & \text{with probability } 1-(z-\lfloor z\rfloor),\\
\lfloor z\rfloor +1 & \text{with probability } z-\lfloor z\rfloor,\\
\end{array}\right\}
\ =\ \lfloor z\rfloor\ +\ \bin(1,\ z-\lfloor z\rfloor).
\ea
Here, $\bin(n,\pi)$ abbreviates the binomial distribution with parameters $n\in\bbn = \{1,2,\ldots\}$ and $\pi\in (0,1)$, see \citet{johnson05}.
The operator $\langle \cdot \rangle$ was referred to as ``first-order random rounding'' by \citet{liu13} and as ``stochastic rounding'' by \citet{gupta15}.
The last equality in \eqref{rounding_operator} shows that the real number~$z$ is split into its integer part~$\lfloor z\rfloor$ and fractional part $z-\lfloor z\rfloor$, in analogy to the so-called ``binomial multiplicative operator'' in \citet{weisszhu24}. As the common notation ``$\{z\}$'' for the fractional part is easily confused with the one for a one-point set (singleton), we use the abbreviation $\widetilde{z}:=z-\lfloor z\rfloor$ instead. The distribution in \eqref{rounding_operator} can also be summarized in terms of its probability generating function (pgf), then
\ba
\label{rounding_operator_pgf}
\textstyle
\pgf_{\langle z\rangle}(s)\ =\ s^{\lfloor z\rfloor}\cdot \e\big[s^{\bin(1, \widetilde{z})}\big]\ =\ (1-\widetilde{z})\,s^{\lfloor z\rfloor} + \widetilde{z}\,s^{\lfloor z\rfloor+1}.
\ea
The last representation in \eqref{rounding_operator} immediately implies that
\ba
\label{rounding_operator_mean}
\e\big[\langle z\rangle\big]
\ =\ \lfloor z\rfloor\ +\ (z-\lfloor z\rfloor)\ =\ z,
\ea
so the mean is equal to~$z$, which implies the mean-preserving property mentioned before: if~$Z$ is a real-valued random variable, then $\e\big[\langle Z\rangle\ |\ Z\big] = Z$, such that the law of total expectation implies that $\e\big[\langle Z\rangle\big] = \e[Z]$. This property was also highlighted by \citet[p.~1739]{gupta15}, who characterized $\langle\cdot\rangle$ as an ``unbiased rounding scheme'', as well as by \citet{liu13}.
\medskip

Using the last representation in \eqref{rounding_operator}, we also get the variance as
\ba
\label{rounding_operator_variance}
\V\big[\langle z\rangle\big]
\ =\
0\ +\ \V\big[\bin(1,\ \widetilde{z})\big]
\ =\
\widetilde{z}\,(1-\widetilde{z}).
\ea
As a consequence, if~$Z$ is a real-valued random variable, then
$\V\big[\langle Z\rangle\ |\ Z\big] = \widetilde{Z}\,\big(1-\widetilde{Z}\big)$. By the law of total variance, we get
\begin{align*}
\V\big[\langle Z\rangle\big]
\ &=\ \V\big[\e[\langle Z\rangle\ |\ Z]\big] + \e\big[\V[\langle Z\rangle\ |\ Z]\big]
\ =\ \V[Z] + \e\big[\widetilde{Z}\,\big(1-\widetilde{Z}\big)\big].
\end{align*}
While a simple closed-form expression for the last term is missing, it is obviously from the interval $[0, 0.25]$, \ie $\V\big[\langle Z\rangle]$ deviates from $\V[Z]$ by at most~0.25.
\medskip

Note that $\langle z\rangle$ can be equal to some $y\in\bbz$ iff $z\in (y-1,y+1)$ holds.
Analogously, for given~$z\in\bbr$, it also holds that $y\in (z-1,z+1)$.
As a consequence, if $x\in\bbz$, then
\ba
\label{xy_range}
\langle \alpha\cdot x\rangle = y
\quad \text{possible iff}\quad
\left\{\begin{array}{ll}
x\in \Big\{\lfloor\tfrac{y-1}{\alpha}\rfloor+1,\ldots,\lceil\tfrac{y+1}{\alpha}\rceil-1\Big\} & \text{if } \alpha\in (0,1),\\[2ex]
x\in \Big\{\lfloor\tfrac{y+1}{\alpha}\rfloor+1,\ldots,\lceil\tfrac{y-1}{\alpha}\rceil-1\Big\} & \text{if } \alpha\in (-1,0).\\
\end{array}\right.
\ea
The case $\alpha=0$ is degenerate as then $\langle 0\cdot x\rangle = 0$ holds without exception.
The set for $\alpha\in (0,1)$ in \eqref{xy_range} can be further split by considering the cases $\alpha x\geq y\ \Leftrightarrow\ x\geq \lceil\tfrac{y}{\alpha}\rceil$ and $\alpha x< y\ \Leftrightarrow\ x\leq \lceil\tfrac{y}{\alpha}\rceil-1$. This distinction is important in view of the definition of the operator $\langle\cdot\rangle$: if $\alpha x\geq y$, then $\langle \alpha\cdot x\rangle = y$ can only happen by selecting $\lfloor \alpha\cdot x\rfloor$, while $\alpha x< y$ requires $\lfloor \alpha\cdot x\rfloor+1$ for obtaining $\langle \alpha\cdot x\rangle = y$.
For $\alpha\in (-1,0)$, an analogous splitting is possible, now between $\alpha x\geq y\ \Leftrightarrow\ x\leq \lfloor\tfrac{y}{\alpha}\rfloor$ and $\alpha x< y\ \Leftrightarrow\ x\geq \lfloor\tfrac{y}{\alpha}\rfloor+1$.
\medskip

Hence, if~$X$ is a $\bbz$-valued random variable, if $\alpha\in (0,1)$, and if $\langle\cdot\rangle$ is executed independently of~$X$, we get
\ba
\label{conv_rounding}
\begin{array}{@{}l}
P\big(\langle \alpha\cdot X\rangle = y\big)
\ =
\sum\limits_{x = \lfloor\tfrac{y-1}{\alpha}\rfloor+1}^{\lceil\tfrac{y+1}{\alpha}\rceil-1} P\big(\langle \alpha\cdot x\rangle = y\big)\cdot P(X=x)
\\[2ex]
\ =
\sum\limits_{x = \lfloor\tfrac{y-1}{\alpha}\rfloor+1}^{\lceil\tfrac{y}{\alpha}\rceil-1} P\big(\langle \alpha\cdot x\rangle = y\big)\cdot P(X=x)
\ +
\sum\limits_{x = \lceil\tfrac{y}{\alpha}\rceil}^{\lceil\tfrac{y+1}{\alpha}\rceil-1} P\big(\langle \alpha\cdot x\rangle = y\big)\cdot P(X=x)
\\[2ex]
\ =
\sum\limits_{x = \lfloor\tfrac{y-1}{\alpha}\rfloor+1}^{\lceil\tfrac{y}{\alpha}\rceil-1} \widetilde{\alpha x}\cdot P(X=x)
\ +
\sum\limits_{x = \lceil\tfrac{y}{\alpha}\rceil}^{\lceil\tfrac{y+1}{\alpha}\rceil-1} \big(1-\widetilde{\alpha x}\big)\cdot P(X=x),
\end{array}
\ea
which is a convolution between two discrete distributions.
Analogously, if $\alpha\in (-1,0)$, then
\ba
\label{conv_rounding_neg}
\begin{array}{@{}l}
P\big(\langle \alpha\cdot X\rangle = y\big)
\ =
\sum\limits_{x = \lfloor\tfrac{y+1}{\alpha}\rfloor+1}^{\lfloor\tfrac{y}{\alpha}\rfloor} \big(1-\widetilde{\alpha x}\big)\cdot P(X=x)
\ +
\sum\limits_{x = \lfloor\tfrac{y}{\alpha}\rfloor+1}^{\lceil\tfrac{y-1}{\alpha}\rceil-1} \widetilde{\alpha x}\cdot P(X=x).
\end{array}
\ea
For $\alpha=0$, we certainly get $P\big(\langle 0\cdot X\rangle = y\big) = \delta_{0y}$, where $\delta_{ab}$ denotes the Kronecker delta, which equals~1 (0) if $a=b$ ($a\not=b$).

\section{A Novel Class of $\bbz$-valued ARMA Models}
\label{A Novel Class of Integer-valued ARMA Models}

\subsection{Model Definition}
\label{Model Definition}
In what follows, we present the main contribution of our article, namely a novel ARMA-like model for $\bbz$-valued time series using the mean-preserving rounding operator $\langle\cdot\rangle$ from \eqref{rounding_operator_mean}, referred to as mean-preserving rounded ARMA (MRARMA) model. There are several ways how such integer-valued MRARMA models can be defined. In our opinion, the most convenient approach is to define the MRARMA$(p,q)$ model by the recursive scheme
\ba
\label{Kachour_type_Definition}
X_t\ =\ \epsilon_t + \big\langle Z_{t-1}\big\rangle
\quad\text{with } Z_{t-1}\ =\ \alpha_1\cdot X_{t-1}+\ldots+\alpha_p\cdot X_{t-p} + \beta_1\cdot\epsilon_{t-1} +\ldots + \beta_q\cdot\epsilon_{t-q},
\ea
where $(\epsilon_t)$ constitutes a sequence of independent and identically distributed (\iid) $\bbz$-valued innovations, and where the operator $\langle\cdot\rangle$ at each time~$t$ is executed independently of the innovations and the available process history.
The parameters $\alpha_1, \ldots, \beta_q\in\bbr$ have to be chosen appropriately, see Theorem~\ref{existence} below.
Note that~$Z_{t-1}$ only depends on information being available up to time~$t-1$, although the innovations are usually not observed in practice. The model order $(p,q)\in\bbn_0^2$ covers the pure mean-preserving rounded autoregressive (MRAR) model (if $q=0$) or pure mean-preserving rounded moving average (MRMA) model (if $p=0$) as special cases, while $p=q=0$ corresponds to \iid\ integers. Comparing \eqref{Kachour_type_Definition} to the existing literature on integer time series models, it becomes clear that this model definition is inspired by the rounded autoregressive (RAR) models of \citet{kachour09,kachour14}, more precisely by the model formulation in \citet{kachour14} as this version avoids the possible identifiability problems of \citet{kachour09}. The major difference of \eqref{Kachour_type_Definition} to \citet{kachour14} is given by the use of the mean-preserving rounding operator $\langle\cdot\rangle$ from \eqref{rounding_operator} instead of the ordinary rounding operator, as this shall allow us to derive closed-form expressions for marginal moments and acf.

\begin{bem}
\label{remarkLiu13}
Our novel MRARMA model \eqref{Kachour_type_Definition} is also related to the ``random rounded INARCH (RRINARCH) model'' (1.7) in \citet{liu13}, where mean-preserving rounding is used as well. However, there are also important differences between our model and the one of \citet{liu13}. First, the RRINARCH model does not have an MA-part but an ARCH-type feedback term instead. Second, even in the pure AR-case ($q=0$ in \eqref{Kachour_type_Definition} and $h_t=1$ in model (1.7) of \citet{liu13}, respectively), both model approaches differ as \citet{liu13} restrict to centered innovations while including a real-valued mean parameter within the rounding operator. The latter construction is analogous to the model of \citet{kachour09}, but it may cause identifiability problems as mentioned before. Such problems are avoided by our model formulation \eqref{Kachour_type_Definition}.
\end{bem}
As indicated before \eqref{Kachour_type_Definition}, there would be further options for defining an ARMA-type model for integer time series by using the operator $\langle\cdot\rangle$ from \eqref{rounding_operator}. One particularly obvious approach is later discussed in Appendix~\ref{On an Alternative Model Definition}. But as we shall recognize there, this alternative ARMA-type model has more sophisticated stochastic properties. Therefore, we focus on definition \eqref{Kachour_type_Definition} in the sequel.

\begin{bem}
\label{remarkCounts}
At first glance, it appears that the MRARMA model \eqref{Kachour_type_Definition} (and also model \eqref{INARMA_type_Definition} from Appendix~\ref{On an Alternative Model Definition}) could also be useful for modeling count time series, which could be achieved by requiring the innovations $(\epsilon_t)$ to be count random variables and the parameters $\alpha_1,\ldots,\beta_q$ to be non-negative. The sample paths being generated by such a recursive scheme, however, are restricted by a limited possibility for downward movements. Given the values of $X_{t-1},\ldots,\epsilon_{t-q}$, the term $\big\langle Z_{t-1}\big\rangle$ in \eqref{Kachour_type_Definition}
will often be truly positive, because downward movements beyond the floor operation are not possible. Thus, transitions towards zero will often be impossible as $\epsilon_t\geq 0$.
\end{bem}

\subsection{Existence of Stationary Solution}
\label{Existence of Stationary Solution}
To establish the existence and stationarity of the MRARMA$(p,q)$ process \eqref{Kachour_type_Definition}, we define the following $(p+q)$-dimensional vectorized process
$$
\fY_t\ :=\ \left(\begin{array}{c}X_t\\X_{t-1}\\ \vdots\\X_{t-p+1}\\\epsilon_t\\\epsilon_{t-1}\\\vdots\\\epsilon_{t-q+1}\end{array}\right)
\ =\
\left(\begin{array}{c} \big\langle Z_{t-1}\big\rangle + \epsilon_t \\X_{t-1}\\ \vdots\\X_{t-p+1}\\\epsilon_t\\\epsilon_{t-1}\\\vdots\\\epsilon_{t-q+1}\end{array}\right).
$$
Then, the process $(\fY_t)$ denotes a homogeneous Markov chain with state space $E^\ast=\mathbb{Z}^{p+q}$ and transition probabilities
$$
\textstyle
\pi(\fx,\fy)=P\left(\epsilon_1=y_1-\Big\langle \sum\limits_{i=1}^p\alpha_i x_i+\sum\limits_{j=1}^q\beta_j x_{p+j}\Big\rangle \right)\,\indfkt_{\{y_2=x_1,\ldots,y_{p}=x_{p-1},y_{p+2}=x_{p+1},\ldots,y_{p+q}=x_{p+q-1}\}},
$$
where $\fx=(x_1,\ldots,x_{p+q})^\top\in E^\ast$ and $\fy=(y_1,\ldots,y_{p+q})^\top\in E^\ast$, and where $\indfkt_{\cdot}$ denotes the indicator function.
Then, the process \eqref{Kachour_type_Definition} can be written in the following form:
$$
\fY_t\ =\ \langle \mA\,\fY_{t-1}\rangle+\feta_t,
$$
where rounding is done component-wise, $\langle \fx\rangle=\left(\langle x_1\rangle,\ldots,\langle x_{p+q}\rangle\right)^\top$, where the $(p+q)\times (p+q)$-matrix $\mA=(a_{kl})$ is defined as
\begin{align*}
  &a_{1i}=\alpha_i,i=1,\ldots,p;~~~ a_{1,j+p}=\beta_j,j=1,\ldots,q;\\
  &a_{i+1,i}=1,i\neq p,i=1,2,\ldots,p+q-1;~~~ a_{kl}=0,~\hbox{otherwise},
\end{align*}
and where $\feta_t=(\epsilon_t,0,\ldots,0,\epsilon_t,0,\ldots,0)^\top$, \ie with $\epsilon_t$ being the value of the first and $(p+1)$-th coordinates, and zero being the value of the remaining coordinates.
\medskip

The following theorem gives the conditions that ensure the ergodicity and
the stationarity of the process \eqref{Kachour_type_Definition}.
For $\fx=(x_1,\ldots,x_{p+q})^\top$,  let $\|\fx\|_1=\sum_{i=1}^{p+q}|x_i|$ denote the $L_1$-norm. In addition, let $\rho(\mA)$ denote the spectral radius of $\mA$. For any measure $\lambda$ and function $g$ on $E^\ast$, we set $\lambda(g)=\int g(x)\,d\lambda(x)$.

\begin{thm}
\label{existence}
  Suppose that
(i)  the probability law of the \iid\ noise $\epsilon_t$ charges all points of $E^\ast$, \ie $P(\epsilon_t=a)>0$ for any $a\in\mathbb{Z}$;
(ii) $E|\epsilon_t|^k<\infty$ for some $k\geq 1$;
(iii) $\alpha_p\neq0,\beta_q\neq0,\rho(\mA)<1$.
Then:
\begin{itemize}
	\item[(1)] The process \eqref{Kachour_type_Definition} $(X_t)$ has an unique invariant probability measure $\lambda$ which has a moment of order $k$, \ie $\lambda(\|\cdot\|_1^k)<\infty$.
	\item[(2)] For any $\fy\in E^\ast$ and $f\in L^1(\lambda)$, we have
$$
\frac{1}{n}\sum_{k=1}^n f(\fY_k)\rightarrow\lambda(f),~~~~P_{\fyi}~a.s.,
$$
where $P_{\fyi}$ denotes the conditional probability $P(\cdot|\fY_0=\fy)$.
\end{itemize}
\end{thm}
The proof of Theorem~\ref{existence} is provided by Appendix~\ref{Proofs}.

\begin{bem}
Assumption (i) is just used to show the irreducibility of $(X_t)$.
Let the left upper corner $p\times p$-submatrix of $\mA$ be $\mA^\ast$, \ie $\mA^\ast=(a_{kl})_{1\leq k,l\leq p}$.
According to \citet{dion95}, we know that $\rho(\mA)=\rho(\mA^\ast)$.
Compared to \citet{kachour14}, although we have different models, rounding operators, and coefficient matrix $\mA$, the assumptions and conclusions in Theorem~\ref{existence} and Theorem 1 in \citet{kachour14} are comparable.
\end{bem}

\begin{bem}
\citet{dion95} provided  conditions for the existence of a stationary distribution for thinning-based generalized INARMA process of general orders based on the theory of multi-type branching processes with immigration, whose results are similar to Theorem \ref{existence}, \ie $\rho(\mM)<1$, where $\mM$ is  the mean matrix of the offspring distribution and $\mM$ is the transpose of $\mA$ considered here. This condition can be further simplified to $0<\sum_{i=1}^p\alpha_i<1$ if the  autoregressive coefficients are positive, which are the same for the existence of a stationary distribution for the thinning-based generalized integer-valued autoregressive  process without the  moving average terms. This phenomenon also happens for the  real-valued ARMA models, see the Remark on page 1123 of \citet{ling99}.
\end{bem}

For thinning-based INARMA process,  \cite{neal07} also assumed that $\sum_{j=1}^p\beta_j<1$.  For a real-valued ARMA process, this condition is sufficient for the time series to be invertible. But it is an open question whether this condition is sufficient for the INARMA process to be invertible. Here, we also make a similar assumption: we require that  $\beta=(\beta_1,\ldots,\beta_q)^\top$ are such that the roots of the $q$th-order polynomial
$$
1-\beta_1x-\beta_2x^2-\ldots-\beta_qx^q
$$
are outside the unit circle.

\subsection{Moments and Probabilistic Properties}
\label{Moments and Probabilistic Properties}
As the operator $\langle\cdot\rangle$ from \eqref{rounding_operator} is mean-preserving, recall \eqref{rounding_operator_mean}, model definition \eqref{Kachour_type_Definition} leads to the linear conditional mean
\ba
\label{CMean}
\e[X_t\ |\ X_{t-1},\ldots,\ \epsilon_{t-1},\ldots]\ =\ \mu_\epsilon + \alpha_1\cdot X_{t-1}+\ldots+\alpha_p\cdot X_{t-p} + \beta_1\cdot\epsilon_{t-1} +\ldots + \beta_q\cdot\epsilon_{t-q},
\ea
where $\mu_\epsilon=\e[\epsilon_t]$. Such a closed-form expression does not exist for the models of \citet{kachour09,kachour14}. It was one of the main motivations for our model proposal(s) as it implies a set of Yule--Walker equations for the acf, see equations \eqref{Cov} and \eqref{Cov2} below. Furthermore, the unconditional mean $\mu=\e[X_t]$ takes the same form as for ordinary ARMA models, namely
\ba
\label{Mean}
\mu\ =\ \mu_\epsilon\,\frac{1 + \beta_1 +\ldots + \beta_q}{1 - \alpha_1-\ldots-\alpha_p},
\ea
where $\alpha_1+\ldots+\alpha_p\ <1$ is required.
\medskip

The autocovariance at lag~$h\in\bbn$, $\gamma(h) := \Cov[X_t, X_{t-h}]$, is computed by using the law of total covariance:
\ba
\label{Cov}
\begin{array}{rl}
\gamma(h)\ =&
\Cov\big[\e[X_t\ |\ X_{t-1},\ldots,\ \epsilon_{t-1},\ldots],\ X_{t-h}\big]
\\[1ex]
=&
\sum_{i=1}^p \alpha_i\,\gamma(h-i) + \sum_{j=1}^q \beta_j\,c(h-j),
\end{array}
\ea
where the mixed covariances $c(s):= \Cov[X_r, \epsilon_{r-s}]$, $s\in\bbn_0$, satisfy $c(0)=\Cov[X_r, \epsilon_{r}] = \V[\epsilon_r] =: \sigma_\epsilon^2$,
while for $s\in\bbn$, we get
\ba
\label{Cov2}
\begin{array}{rl}
c(s) = \Cov[X_r, \epsilon_{r-s}]\ =&
\Cov\big[\e[X_r\ |\ X_{r-1},\ldots,\ \epsilon_{r-1},\ldots],\ \epsilon_{r-s}\big]
\\
=&
\sum_{i=1}^p \alpha_i\,\Cov[X_{r-i}, \epsilon_{r-s}] + \sum_{j=1}^q \beta_j\,\Cov[\epsilon_{r-j}, \epsilon_{r-s}]
\\
=&
\sum_{i=1}^{\min\{s,p\}} \alpha_i\,c(s-i) + \sum_{j=1}^q \beta_j\,\delta_{js}\,\sigma_\epsilon^2.
\end{array}
\ea
Equations \eqref{Cov} and \eqref{Cov2} just correspond to the Yule--Walker equations of an ordinary ARMA model, \ie the MRARMA model \eqref{Kachour_type_Definition} has the same autocorrelation structure as an ordinary ARMA model.

\begin{figure}[t]
\centering\footnotesize
\includegraphics[viewport=0 45 550 230, clip=, scale=0.7]{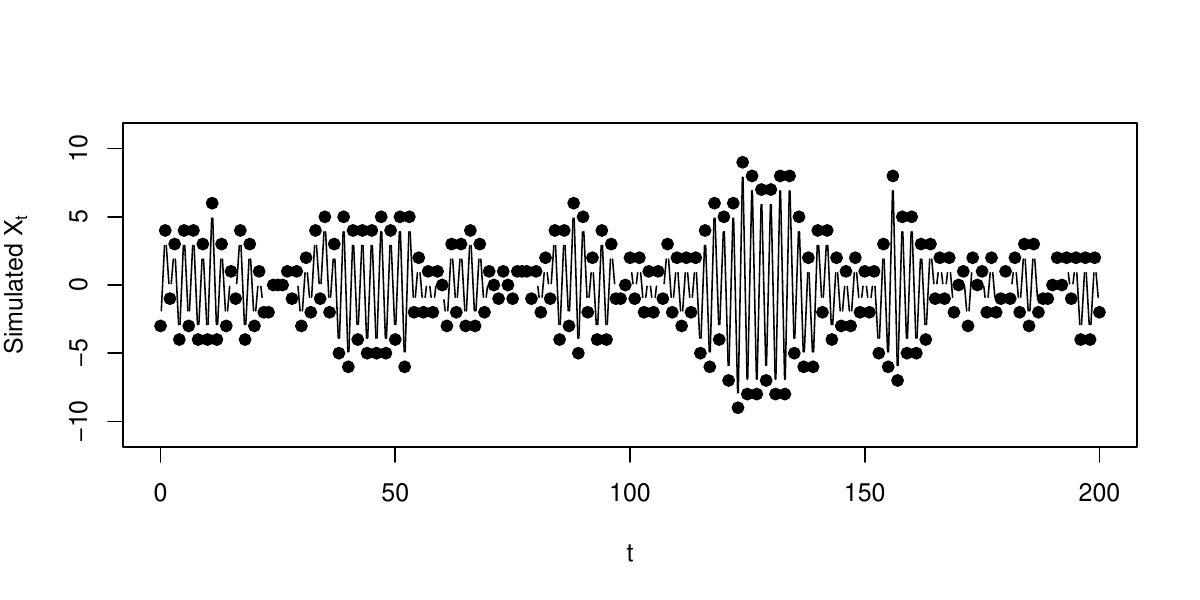}$t$
\caption{Simulated sample path of MRAR$(2)$ model \eqref{Kachour_type_Definition} with $\alpha_1=-0.6$, $\alpha_2=0.3$, and Skellam innovations with $(\mu_\epsilon,\sigma_\epsilon^2)=(0,2)$.}
\label{figSimRRAR2}
\end{figure}
\medskip

Let us now turn to the (conditional) variance of the MRARMA model \eqref{Kachour_type_Definition}. Here, \eqref{rounding_operator_variance} implies the simple expression
\ba
\label{CVar_Kachour}
\V[X_t\ |\ X_{t-1},\ldots,\ \epsilon_{t-1},\ldots]\ =\ \sigma_\epsilon^2 + \widetilde{Z_{t-1}}\,\big(1-\widetilde{Z_{t-1}}\big),
\ea
which can be used in practice for computing, \eg the standardized Pearson residuals as a popular tool for model diagnostics \citep[see][Section~2.4]{weiss18}.
Since $\widetilde{Z_{t-1}}\,\big(1-\widetilde{Z_{t-1}}\big)$ takes a value from $[0,0.25]$, we have relatively strong heteroscedasticity for low~$\sigma_\epsilon^2$ and vice versa.
An illustrative simulated sample path is shown in Figure~\ref{figSimRRAR2}, where Skellam-distributed innovations \citep[see][Section~4.12.3]{johnson05} are used.
The unconditional variance of the MRARMA model \eqref{Kachour_type_Definition} satisfies
\ba
\label{Var_Kachour}
\begin{array}{rl}
\sigma^2 := \V[X_t]\ =& \V\big[\e[X_t\ |\ X_{t-1},\ldots,\ \epsilon_{t-1},\ldots]\big] + \e\big[\V[X_t\ |\ X_{t-1},\ldots,\ \epsilon_{t-1},\ldots]\big]
\\
=&
 \V[Z_{t-1}] + \sigma_\epsilon^2 +  \e\big[\widetilde{Z_{t-1}}\, \big(1-\widetilde{Z_{t-1}}\big)\Big].
\end{array}
\ea
Here, the first summand $\V[Z_{t-1}] = \V[\alpha_1\cdot X_{t-1}+\ldots+\alpha_p\cdot X_{t-p} + \beta_1\cdot\epsilon_{t-1} +\ldots + \beta_q\cdot\epsilon_{t-q}]$ requires to evaluate the Yule--Walker equations \eqref{Cov} and \eqref{Cov2} for the autocovariance function, whereas the last term $\e\big[\widetilde{Z_{t-1}}\, \big(1-\widetilde{Z_{t-1}}\big)\big]$ takes a value from the interval $[0, 0.25]$. Exactly if~$Z_{t-1}$ is integer-valued, the last term becomes zero as for ordinary ARMA models, but the possible deviation from ordinary ARMA models is bounded from above by~0.25.

\begin{figure}[t]
\centering\scriptsize
(a)\hspace{-3ex}\includegraphics[viewport=0 45 335 305, clip=, scale=0.6]{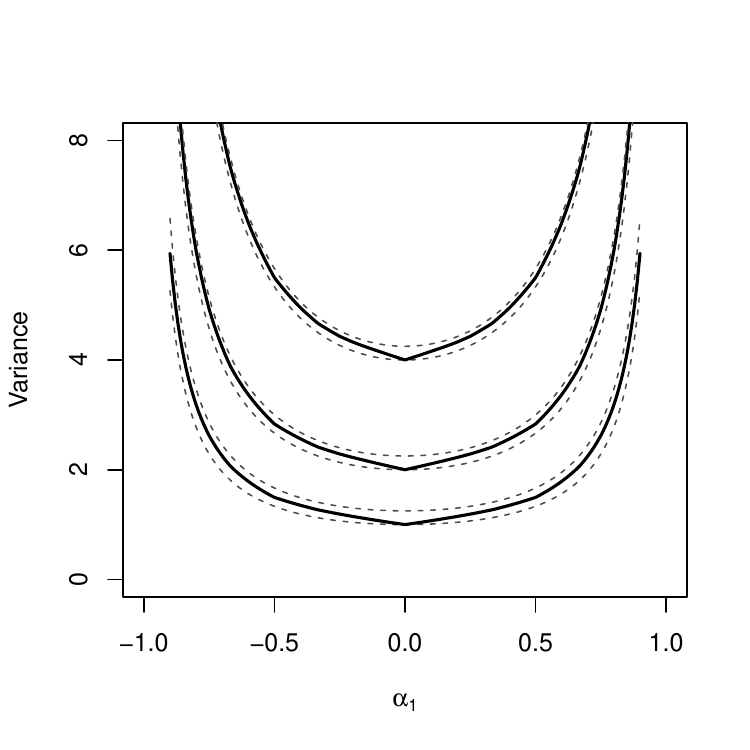}$\alpha_1$
\qquad
(b)\hspace{-3ex}\includegraphics[viewport=0 45 335 305, clip=, scale=0.6]{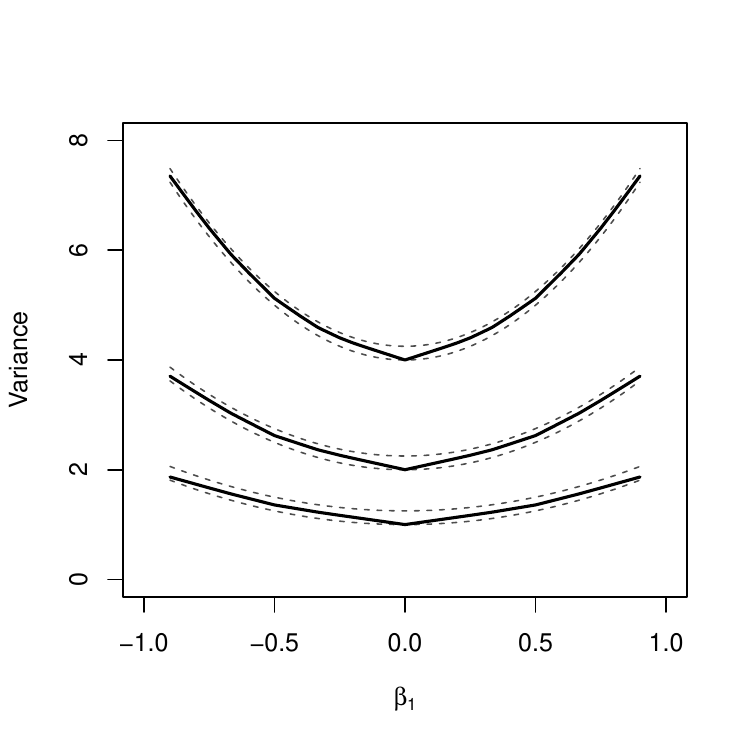}$\beta_1$
\caption{Example~\ref{exampleVar}: True variances (black solid lines) and variance bounds (grey dashed lines) for (a) MRAR$(1)$ process against~$\alpha_1$ and (b) MRMA$(1)$ process against~$\alpha_1$. Innovations from Skellam distribution with mean~0 and variances~1, 2, 4 (lowest graphs to uppermost graphs).}
\label{figVar}
\end{figure}

\begin{bsp}
\label{exampleVar}
In the special case of an MRAR$(1)$ model,
formula \eqref{Var_Kachour} leads to
$$
\sigma^2\,(1-\alpha_1^2)\ =\ \sigma_\epsilon^2 + \e\big[\widetilde{Z_{t-1}}\, \big(1-\widetilde{Z_{t-1}}\big)\Big]\quad \in\ \sigma_\epsilon^2 + [0,0.25].
$$
These variance bounds are illustrated in Figure~\ref{figVar}\,(a) by dashed lines for Skellam-distributed innovations. Also the true variance is shown as a solid line, which is computed from the stationary marginal distribution, see Example~\ref{exampleProbs} for details. It can be recognized that the variance bounds are narrow and close to the true variance, \ie they provide a good approximation of the true variance.
\medskip

For the MRMA$(q)$ model according to \eqref{Kachour_type_Definition}, in turn,
$$
\textstyle
\V[Z_{t-1}] = \sum_{i,j=1}^q \Cov[\beta_i\cdot\epsilon_{t-i}, \beta_j\cdot\epsilon_{t-j}] = \sum_{j=1}^q \beta_j^2\,\sigma_\epsilon^2,
$$
so $\sigma^2\, =\, \sigma_\epsilon^2\,(1 + \sum_{j=1}^q \beta_j^2) + \e\big[\widetilde{Z_{t-1}}\, \big(1-\widetilde{Z_{t-1}}\big)\big]$, which is the ordinary MA$(q)$'s variance plus a value from $[0,0.25]$. An illustrative plot for the MRMA$(1)$-type model is shown in Figure~\ref{figVar}\,(b). Here, the true variance is again computed from the stationary marginal distribution, see Example~\ref{exampleProbs}. Like before, the variance bounds provide a close approximation to the true variance.
\end{bsp}

While the variance bounds in Example~\ref{exampleVar} only make use of the innovations' variance~$\sigma_\epsilon^2$, the true variance~$\sigma^2$ is also (slightly) affected by further distributional properties such as the innovations' mean, see Example~\ref{exampleProbs} below; these (minor) effects are covered by the term $\e\big[\widetilde{Z_{t-1}}\, \big(1-\widetilde{Z_{t-1}}\big)\big]$.
\medskip

For purely autoregressive models, also conditional distributions can be derived. For the MRAR$(p)$ model according to \eqref{Kachour_type_Definition}, we have the convolution
$$
\textstyle
P(X_t=x\ |\ X_{t-1}=x_1,\ldots)\ =\ \sum\limits_{k=-\infty}^\infty P(\epsilon_t=x-k)\,  P\big(\langle z\rangle=k\big)\
\quad\text{with } z=\alpha_1 x_1 +\ldots+ \alpha_p x_p.
$$
But as we know from the discussion before \eqref{xy_range}, $\langle z\rangle=k$ is possible only if $k\in (z-1,z+1)\cap\bbz$: if $z\in\bbz$ then $k=z$ has to hold, otherwise $k\in\big\{\lfloor z\rfloor, \lfloor z\rfloor+1\big\}$. Therefore, considering the probabilities in \eqref{rounding_operator}, we obtain
\ba
\label{transprob_Kachour}
\begin{array}{rl}
P(X_t=x\ |\ X_{t-1}=x_1,\ldots)\ =&
(1-\widetilde{z})\, P(\epsilon_t=x-\lfloor z\rfloor) + \widetilde{z}\, P(\epsilon_t=x-\lfloor z\rfloor-1)
\\[1ex]
& \text{with } z=\alpha_1 x_1 +\ldots+ \alpha_p x_p,
\end{array}
\ea
which also covers the case $z\in\bbz$ as then $\lfloor z\rfloor=z$ and $\widetilde{z}=0$. This conditional distribution can also be expressed in terms of the pgf as
\ba
\label{condpgf_Kachour}
\e\big[s^{X_t}\ |\ X_{t-1}=x_1,\ldots\big]\ =\ \pgf_\epsilon(s)\cdot \pgf_{\langle \alpha_1 x_1 +\ldots+ \alpha_p x_p\rangle}(s),
\ea
where $\pgf_\epsilon(s) = \e[s^{\epsilon_t}]$ denotes the innovations' pgf, and where the second factor is computed via \eqref{rounding_operator_pgf}.

\begin{figure}[t]
\centering\footnotesize
(a)\hspace{-3ex}\includegraphics[viewport=0 45 260 235, clip=, scale=0.5]{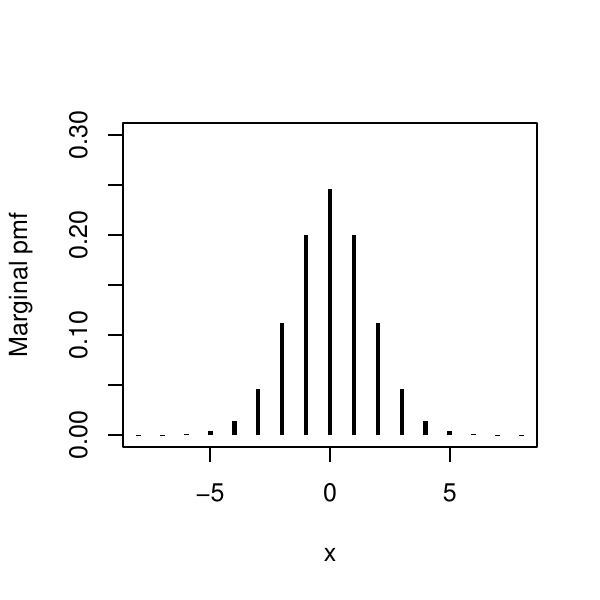}\hspace{-1ex}$x$
\qquad
(b)\hspace{-3ex}\includegraphics[viewport=0 45 260 235, clip=, scale=0.5]{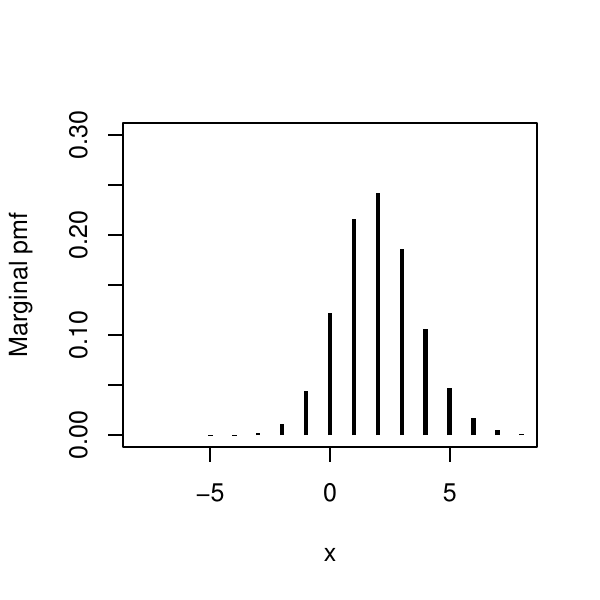}\hspace{-1ex}$x$
\qquad
(c)\hspace{-3ex}\includegraphics[viewport=0 45 260 235, clip=, scale=0.5]{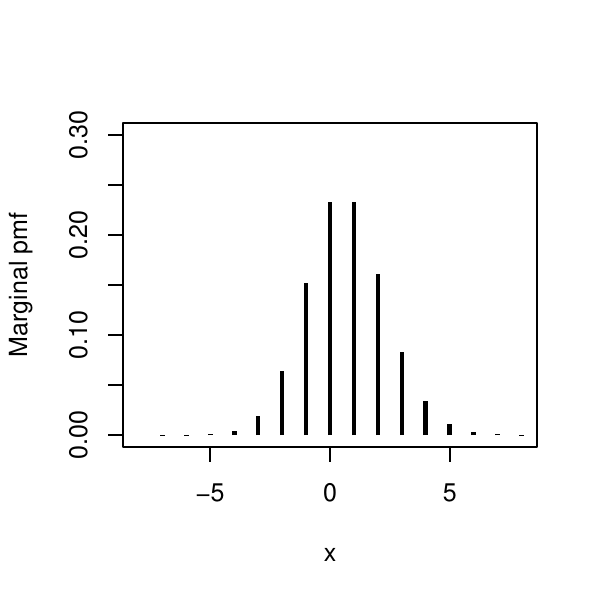}\hspace{-1ex}$x$
\caption{Example~\ref{exampleProbs}: Plots of marginal pmf $P(X_t=x)$ against~$x$ for MRAR$(1)$ model with Skellam innovations, with $(\alpha_1,\mu_\epsilon,\sigma_\epsilon^2)$ equal to (a) $(\pm 0.5, 0, 2)$, (b) $(0.5, 1, 2)$, and (c) $(-0.5, 1, 2)$.}
\label{figPMF}
\end{figure}

\begin{bsp}
\label{exampleProbs}
For the MRAR$(1)$ model (which constitutes a discrete Markov chain),
the transition probabilities follow from \eqref{transprob_Kachour} as
$$
p(x | y) := P(X_t=x\ |\ X_{t-1}=y)\ =\
(1-\widetilde{\alpha_1 y})\cdot P\big(\epsilon_t=x-\lfloor \alpha_1 y\rfloor\big) + \widetilde{\alpha_1 y}\cdot P\big(\epsilon_t=x-\lfloor \alpha_1 y\rfloor-1\big).
$$
The probability mass function (pmf) $P(X_t=x)$ of the MRAR$(1)$'s stationary marginal distribution is computed numerically by solving the corresponding invariance equations, see Remark~2.1.3.4 in \citet{weiss18} for details. The stationary marginal distribution of the MRMA$(1)$ model, in turn, is directly available by the convolution
$$
\textstyle
P(X_t=x)
\ =\ \sum\limits_{k=-\infty}^\infty P(\epsilon_t=x-k)\cdot  P\big(\langle \beta_1 \epsilon_{t-1}\rangle=k\big),
$$
where the latter probability is computed by \eqref{conv_rounding} and \eqref{conv_rounding_neg}. In the previous Example~\ref{exampleVar}, we used the numerically computed stationary marginal distributions to compute the true variance.

\smallskip
Illustrative plots of the marginal pmf of MRAR$(1)$ models with Skellam innovations are shown in Figure~\ref{figPMF}. In part~(a), the innovations have a symmetric distribution with $\mu_\epsilon=0$, and the same is true for the marginal pmf with $\sigma^2\approx 2.83318$, irrespective of the sign of~$\alpha_1$.
In parts~(b) and~(c), however, the innovations have an asymmetric distribution with $\mu_\epsilon=1$, and this asymmetry carries over to the marginal pmf. Furthermore, the pmf is now also affected by the sign of~$\alpha_1$. In~(b) with $\alpha_1=0.5$, we have $\mu = 2$ and $\sigma^2\approx 2.83345$, whereas $\mu = 2/3$ and $\sigma^2\approx 2.83320$ for $\alpha_1=-0.5$ in~(c). We provide five digits for~$\sigma^2$ to show that there are slight differences depending on the actual properties of the innovations' distribution. But all variances satisfy the unique variance bounds~$[2.667, 3]$.

\smallskip
The aforementioned symmetry in Figure~\ref{figPMF}\,(a) can also be proven formally by utilizing the following properties of the floor operation and the fractional part: for an arbitrary real number $z\in\bbr$, it holds that $\lfloor -z\rfloor = -\lfloor z\rfloor - \indfkt_{z\not\in\bbz}$ and $\widetilde{-z} = -\widetilde{z} + \indfkt_{z\not\in\bbz}$. Thus, if $\alpha_1 y\not\in\bbz$, we get for the transition probabilities that
$$
\begin{array}{rl}
p(-x | -y)
\ =&
(1-\widetilde{-\alpha_1 y})\cdot P\big(\epsilon_t=-x-\lfloor -\alpha_1 y\rfloor\big) + \widetilde{-\alpha_1 y}\cdot P\big(\epsilon_t=-x-\lfloor -\alpha_1 y\rfloor-1\big)
\\[1ex]
\ =&
\widetilde{\alpha_1 y}\cdot P\big(\epsilon_t=-x+\lfloor \alpha_1 y\rfloor+1\big) + (-\widetilde{\alpha_1 y}+1)\cdot P\big(\epsilon_t=-x+\lfloor \alpha_1 y\rfloor\big)
\\[1ex]
\ =&
(1-\widetilde{\alpha_1 y})\cdot P\big(\epsilon_t=-(x-\lfloor \alpha_1 y\rfloor)\big) + \widetilde{\alpha_1 y}\cdot P\big(\epsilon_t=-(x-\lfloor \alpha_1 y\rfloor-1)\big),
\end{array}
$$
which equals $p(x | y)$ if~$\epsilon_t$ has a symmetric distribution. For $\alpha_1 y\in\bbz$, by contrast, the transition probabilities have a more simple form anyway, $p(x | y) = P\big(\epsilon_t=x-\alpha_1 y\big)$, such that $p(-x | -y) = P\big(\epsilon_t=-(x-\alpha_1 y)\big)$ again equals $p(x | y)$ if~$\epsilon_t$ has a symmetric distribution.
These symmetries imply the symmetry of the marginal distribution.
\end{bsp}

\section{Performance of Parameter Estimation}
\label{Performance of Parameter Estimation}
First, we discuss estimation methods for the general MRARMA model, then focus on the MRAR model.
\medskip

Due to its close relationship to the ordinary ARMA model, the parameters of the MRARMA model can be estimated by analogous estimation approaches.
First, recall that a direct proof for the consistency and asymptotic normality of maximum likelihood estimation (MLE) for  causal and invertible real-valued ARMA was given by \cite{yao06}, which is difficult to be generalized to the general MRARMA model considered here. Now we use three-stage least squares (LS) estimation method in \cite{koreisha90} to estimate regression parameters $(\alpha_1,...,\alpha_p,\beta_1,...,\beta_q)$. At the first stage, an autoregression of order $\lfloor \sqrt n\rfloor$ is fitted to data $X_1,...,X_n$, wherein the autoregressive parameters are estimated using LS. Denote $\hat\epsilon_t$ the estimate of $\epsilon_t$ obtained from the residuals from this AR. At the second stage, consider the following regression model
\begin{equation}\label{reg1}
X_t-\hat\epsilon_t=\sum_{i=1}^p\alpha_iX_{t-i}+\sum_{j=1}^q\beta_j\hat\epsilon_{t-j}+\beta(B)(\epsilon_t-\hat\epsilon_t),
\end{equation}
where $B$ is the backshift operator $B\epsilon_t=\epsilon_{t-1}$, and $\beta(B)=1+\sum_{j=1}^q\beta_jB^j$. The error term is $\beta(B)(\epsilon_t-\hat\epsilon_t)$, and
a linear regression of $X_t-\hat\epsilon_t$ on $X_{t-1},...,X_{t-p},\hat\epsilon_{t-1},...,\hat\epsilon_{t-q}$ gives LS parameter estimates. At the third stage, the above regression \eqref{reg1} is recalculated but now using a generalized LS procedure as if the errors in the regression were generated by a moving average of order $q$ with parameters $\beta_1,...,\beta_q$, and the estimators in the second stage are used to calculate the covariance matrix of the errors.  The above three-stage LS estimation was improved by \cite{kapetanios03}, who suggested to iterate this LS operation using the new estimate of the error sequence until the estimate of the error sequence converges, and the superiority of this method lies in the fact that it eliminates the random noise
component underlying the justification of the generalized LS application.
\medskip

In the following, we will focus on the pure autoregressive case for ease of computations, i.e., the MRAR$(p)$ model.
Using the Yule--Walker equations \eqref{Cov} and \eqref{Cov2} together with \eqref{Mean} for the mean, the method of moments (MM) is easily implement,  whereas \eqref{CMean} can be used to compute conditional LS (CLS) estimates.
The consistency and asymptotic normality of the CLS estimator can be easily established using arguments in \cite{tjo86}.
\medskip

If, in addition, the distribution of the innovations $(\epsilon_t)$ is specified, also MLE is possible by using the conditional probabilities \eqref{transprob_Kachour}. Since no closed-form formulae are available for MLE, the practical computation is done by numerically maximizing the (conditional) log-likelihood function, where MM or CLS estimates can be used as starting values.
 If the time series $X_1,\ldots,X_n$ origins from an MRAR$(p)$ model, and if~$\ftheta$ denotes the vector of all model parameters, then the conditional log-likelihood function equals
$$
\textstyle
\ell_p(\ftheta)\ =\ \sum_{t=p+1}^n \ln{P(X_t\ |\ X_{t-1},\ldots,X_{t-p})},
$$
which is to be maximized over~$\ftheta$. Since our randomized rounding does not cause identifiability issues, we can also compute approximate standard errors (s.\,e.) from the inverse Hessian at the maximum, see \citet[p.~235]{weiss18} for details. Finally, the maximal value of~$\ell_p(\ftheta)$, say~$\ell_{\max,p}$, can be used for model identification, in the case that multiple candidate models are fitted to the time series. This is done by computing an appropriate information criterion, such as Akaike's (AIC) or the Bayesian one (BIC). Here, we follow the recommendation in \citet[p.~236]{weiss18} and account for the conditioning in MLE computation by using the corrective factor $n/(n-p)$, \ie AIC or BIC are computed by using $\frac{n}{n-p}\,\ell_{\max,p}$ as the maximized log-likelihood value.

\begin{figure}[t]
\centering\footnotesize
$\alpha_1$\hspace{-5ex}\includegraphics[viewport=15 15 150 305, clip=, scale=0.7]{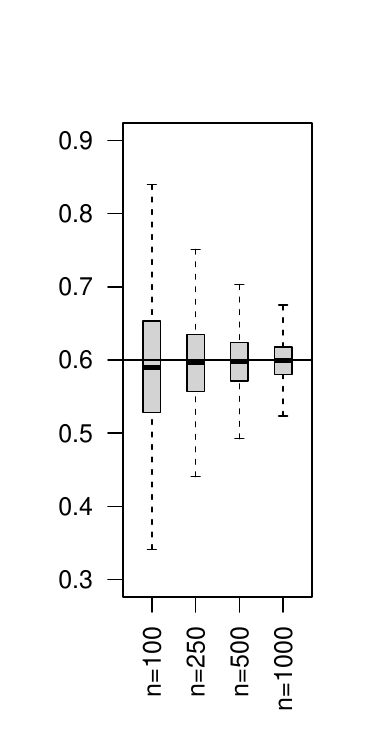}
\qquad\quad
$\alpha_2$\hspace{-5ex}\includegraphics[viewport=15 15 150 305, clip=, scale=0.7]{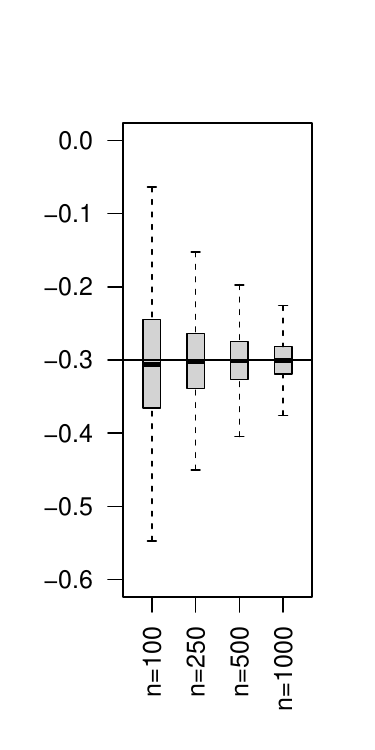}
\qquad\quad
$\lambda_1$\hspace{-5ex}\includegraphics[viewport=15 15 150 305, clip=, scale=0.7]{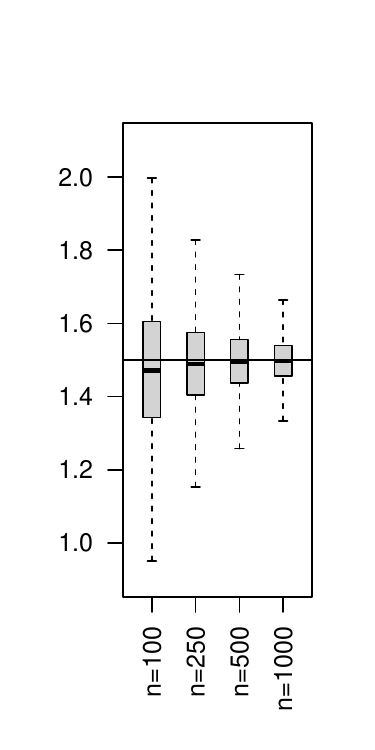}
\qquad\quad
$\lambda_2$\hspace{-5ex}\includegraphics[viewport=15 15 150 305, clip=, scale=0.7]{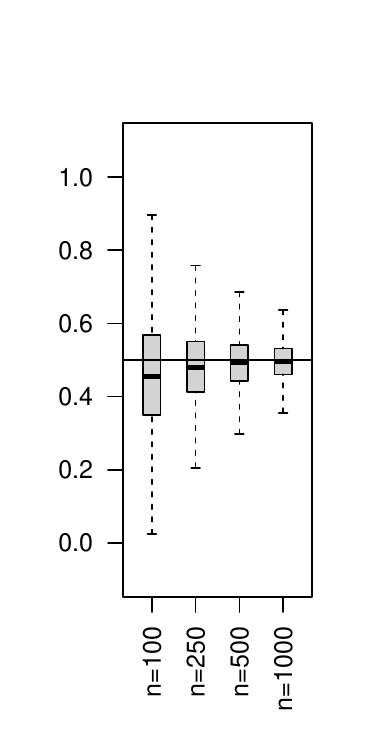}
\caption{Boxplots of MLE corresponding to Skellam-MRAR$(2)$ model with $(\alpha_1,\alpha_2,\lambda_1,\lambda_2)=(0.6,-0.3,1.5,0.5)$.}
\label{figBoxplots}
\end{figure}

\medskip
To analyze the actual performance of the MLE approach, we did a simulation study, where the data-generating process (DGP) is chosen as MRAR$(1)$ or MRAR$(2)$ with Skellam-distributed innovations. Denoting the Skellam parameters by $\lambda_1,\lambda_2>0$, such that $\mu_\epsilon=\lambda_1-\lambda_2$ and $\sigma_\epsilon^2=\lambda_1+\lambda_2$ \citep[see][Section~4.12.3]{johnson05}, the vector of parameters is $\ftheta=(\alpha_1,\lambda_1,\lambda_2)$ or $(\alpha_1,\alpha_2,\lambda_1,\lambda_2)$, respectively. For various parametrizations of the Skellam-MRAR DGPs, we simulated 10,000 time series of length~$n\in\{100,250,500,1000\}$ (plus a pre-run of length~250 for burn-in), which were used for computing parameter estimates and approximate s.\,e.\ as described before. The full simulation results are tabulated in Supplement~S.1; an illustrative example graph with boxplots of ML estimates is shown in Figure~\ref{figBoxplots}. The following conclusions are based on the full simulation results in Supplement~S.1. It can be seen that in any case, the means of estimates quickly converge to the true parameter values for increasing~$n$, and the s.\,e.\ quickly decrease with~$n$, also see the illustrative Figure~\ref{figBoxplots}. This confirms the consistency of the MLE approach. In fact, a certain bias only occurs for the smallest sample size $n=100$, and here either for the AR~parameters in case these have positive values, or for the parameters~$\lambda_1, \lambda_2$ of the Skellam innovations. Let us now compare the approximate s.\,e.\ to the simulated ones (third block of columns vs.\ second one in Supplement~S.1). For the MRAR$(1)$ DGP, we have an excellent agreement in case of~$\lambda_1, \lambda_2$, whereas the approximate s.\,e.\ are slightly too low regarding~$\alpha_1$. For MRAR$(2)$, we now have moderate discrepancies for all parameters (the approximate s.\,e.\ are somewhat too low in the mean), but these discrepancies quickly decrease with increasing~$n$. Thus, for low sample size, the approximate s.\,e.\ should be interpreted with some caution as they possibly (somewhat) underestimate the true s.\,e.

\section{Real-world Data Applications}
\label{Real-world Data Applications}

\begin{figure}[t]
\centering\small
(a)\hspace{-3ex}\includegraphics[viewport=0 45 405 235, clip=, scale=0.65]{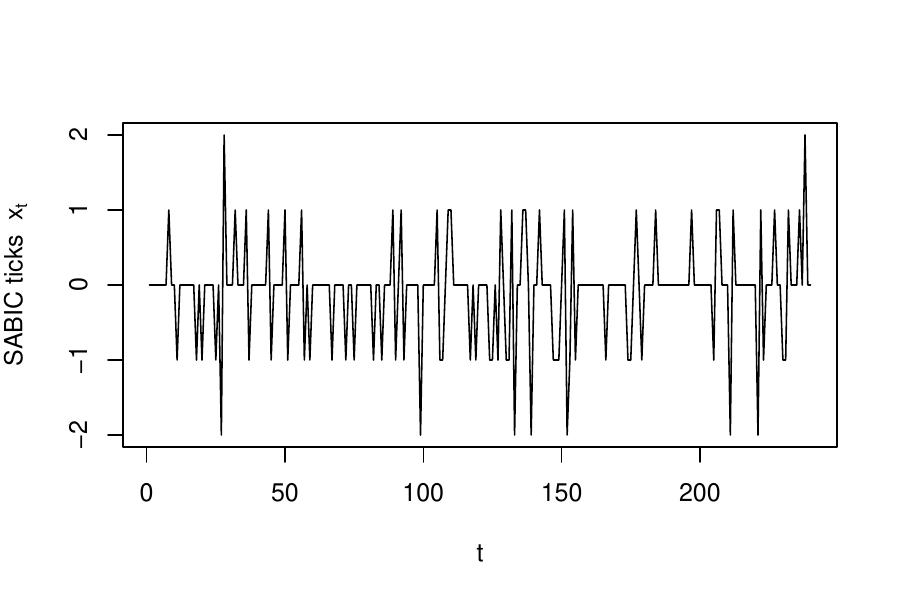}\hspace{-1ex}$t$
\qquad
(b)\hspace{-3ex}\includegraphics[viewport=0 45 190 235, clip=, scale=0.65]{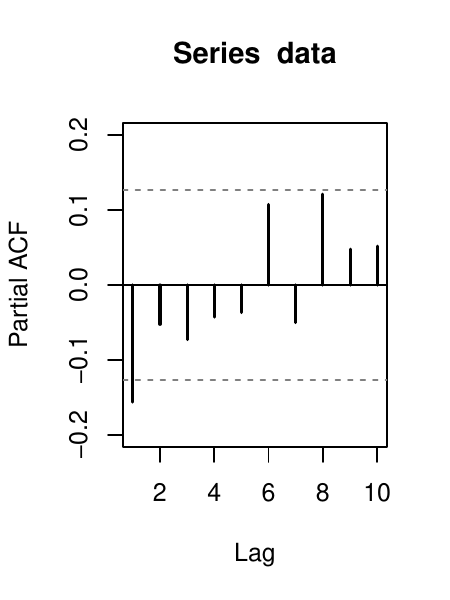}\ $k$
\caption{SABIC ticks data from Section~\ref{Financial Time Series}: (a) time series plot and (b) sample pacf against lag~$k$.}
\label{figSABIC}
\end{figure}

\subsection{Financial Time Series}
\label{Financial Time Series}
When applying our MRARMA models to real-world data examples on $\bbz$-valued time series, it turned out that ticks data from stock exchange are often well-described by low-order MRAR$(p)$ models. As an illustrative example, we shall analyze the Saudi Basic Industry (SABIC) time series from \citet{alzaid10}, but a few further data examples from the literature are briefly presented in Supplement~S.2. The SABIC series consists of (re-scaled) price changes per minute, recorded at June 30, 2007, between 11:15 am and 3:15 pm (thus $n=240$), see Figure~\ref{figSABIC}\,(a). In \citet{alzaid10}, the data were treated as being \iid\ and modelled by a Skellam distribution. The sample partial acf (pacf) in Figure~\ref{figSABIC}\,(b), however, takes a significantly negative value at lag~1, indicating that an AR$(1)$-type model might be more appropriate to describe the data. For this reason, we fitted MRAR$(p)$ models with $p\in\{1,2\}$ and Skellam innovations to the data. As a further competitor, we also consider the RAR$(1)$ model with Skellam innovations by \citet{kachour14}, which differs from our MRAR$(1)$ model by using the ordinary (deterministic) rounding operator instead of the randomized one.

\begin{table}[t]
\centering
\caption{SABIC ticks data from Section~\ref{Financial Time Series}: ML estimates (approximate s.\,e.\ in parentheses) and information criteria for different candidate models.}
\label{tabSABIC}
$$
\begin{array}{l|@{\qquad}rrrr@{\qquad}cc}
\toprule
\text{Skellam-} & \lambda_1 & \lambda_2 & \alpha_1 & \alpha_2 & \text{AIC} & \text{BIC} \\
\midrule
\text{\iid} & 0.168 & 0.251 &  &  & 458.7 & 465.7 \\
 & \text{\footnotesize (0.030)} & \text{\footnotesize (0.036)} &  &  &  &  \\
\midrule
\text{MRAR}(1) & 0.142 & 0.230 & -0.140 &  & 455.1 & 465.5 \\
 & \text{\footnotesize (0.029)} & \text{\footnotesize (0.035)} & \text{\footnotesize (0.062)} &  &  &  \\
\midrule
\text{MRAR}(2) & 0.135 & 0.229 & -0.145 & -0.051 & 457.4 & 471.4 \\
 & \text{\footnotesize (0.030)} & \text{\footnotesize (0.036)} & \text{\footnotesize (0.062)} & \text{\footnotesize (0.058)} &  &  \\
\midrule
\text{RAR}(1) & 0.169 & 0.253 & -0.156 &  & 461.9 & 472.3 \\
 & \text{\footnotesize (---)} & \text{\footnotesize (---)} & \text{\footnotesize (---)} &  &  &  \\
\bottomrule
\end{array}
$$
\end{table}
\medskip

The estimation results (with approximate s.\,e.\ in parentheses) being summarized in Table~\ref{tabSABIC} show that the MRAR$(1)$ model is optimal in terms of both AIC and BIC. It also has significant estimates (on a 5\%-level) throughout, even if we take into account that the true s.\,e.\ might be slightly larger than the approximate ones (recall Section~\ref{Performance of Parameter Estimation}). For the MRAR$(2)$ model, by contrast, the estimate of~$\alpha_2$ is not significant, confirming our above conclusion from the sample pacf that the data are AR$(1)$-like. In this regard, it is interesting to look at the MLE results of the competing RAR$(1)$ model. First, it can be observed that AIC and BIC are clearly worst. Second, we cannot provide approximate s.\,e.\ for this model, because the use of the deterministic rounding operator instead of our randomized mean-preserving one leads to a singular Hessian matrix. Thus, besides the advantages described in Section~\ref{Moments and Probabilistic Properties} (like closed-form moment formulae), a further benefit of using MRAR rather than RAR is the possibility for easily approximating the s.\,e.\ of MLE. Finally, to check the adequacy of the fitted MRAR$(1)$ model, we computed the standardized Pearson residuals \citep[see][Section~2.4]{weiss18}. Their sample acf does not show significant values, their mean $\approx -0.009$ is very close to zero, and their variance $\approx 0.999$ is very close to one. Therefore, the MRAR$(1)$ model is not only the best among the candidate models, it even proved to adequately describe the SABIC ticks data (recall that analogous conclusions also hold for the other financial data sets considered in Supplement~S.2).

\begin{figure}[t]
\centering\small
(a)\hspace{-3ex}\includegraphics[viewport=0 45 405 235, clip=, scale=0.65]{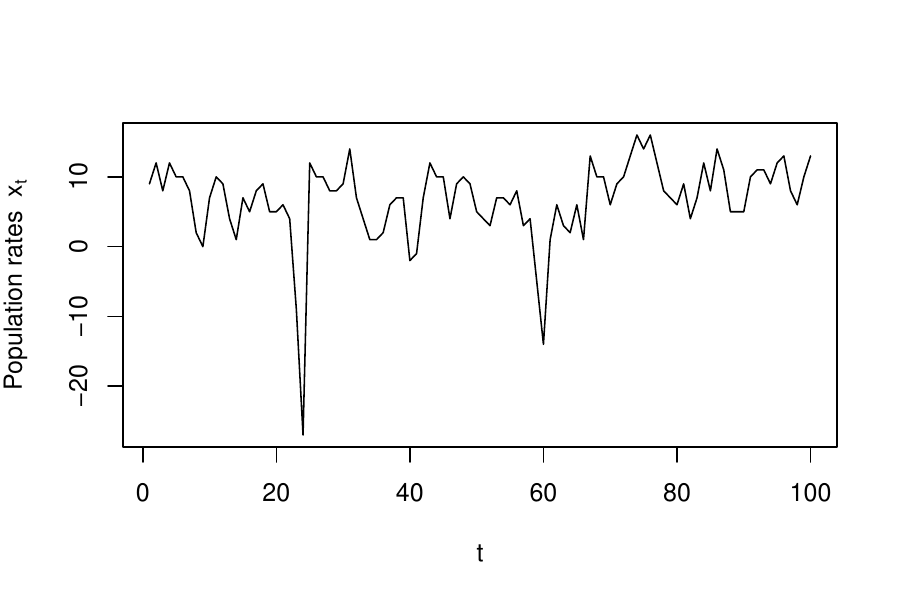}\hspace{-1ex}$t$
\qquad
(b)\hspace{-3ex}\includegraphics[viewport=0 45 190 235, clip=, scale=0.65]{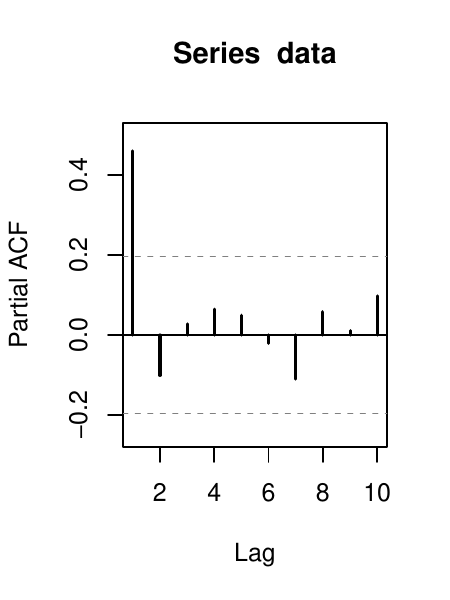}\ $k$
\caption{Swedish population rates data from Section~\ref{Demographic Time Series}: (a) time series plot and (b) sample pacf against lag~$k$.}
\label{figSwedPop}
\end{figure}

\subsection{Demographic Time Series}
\label{Demographic Time Series}
As a second illustrative example, we analyze the Swedish population rates time series from \citet{kachour09}. This time series consists of yearly population rates (per thousand population) between 1750 and 1849 (thus $n=100$), and it is plotted in Figure~\ref{figSwedPop}\,(a). The sample pacf in Figure~\ref{figSwedPop}\,(b) again indicates an AR$(1)$-like autocorrelation structure, but this time with a (rather strong) positive lag-1 value. In view of the similar data properties, we use the same candidate models as in Section~\ref{Financial Time Series}. Here, it should be noted that we still use the (Skellam) RAR$(1)$ model of \citet{kachour14} as a competitor, as the previous model version by \citet{kachour09} suffers from some identifiability problems that avoid from using the MLE approach for model fitting.

\begin{table}[t]
\centering
\caption{Swedish population rates data from Section~\ref{Demographic Time Series}: ML estimates (approximate s.\,e.\ in parentheses) and information criteria for different candidate models.}
\label{tabSwedPop}
$$
\begin{array}{l|@{\qquad}rrrr@{\qquad}cc}
\toprule
\text{Skellam-} & \lambda_1 & \lambda_2 & \alpha_1 & \alpha_2 & \text{AIC} & \text{BIC} \\
\midrule
\text{\iid} & 20.607 & 13.917 &  &  & 641.3 & 646.5 \\
 & \text{\footnotesize (2.453)} & \text{\footnotesize (2.440)} &  &  &  &  \\
\midrule
\text{MRAR}(1) & 14.570 & 11.218 & 0.500 &  & 618.1 & 625.9 \\
 & \text{\footnotesize (1.938)} & \text{\footnotesize (1.930)} & \text{\footnotesize (0.005)} &  &  &  \\
\midrule
\text{MRAR}(2) & 14.864 & 10.995 & 0.493 & -0.077 & 619.8 & 630.2 \\
 & \text{\footnotesize (1.942)} & \text{\footnotesize (1.932)} & \text{\footnotesize (0.061)} & \text{\footnotesize (0.057)} &  &  \\
\midrule
\text{RAR}(1) & 14.899 & 11.334 & 0.460 &  & 618.8 & 626.6 \\
 & \text{\footnotesize (---)} & \text{\footnotesize (---)} & \text{\footnotesize (---)} &  &  &  \\
\bottomrule
\end{array}
$$
\end{table}

\medskip
The estimation results (with approximate s.\,e.\ in parentheses) being summarized in Table~\ref{tabSwedPop} show that the MRAR$(1)$ model is again optimal in terms of both AIC and BIC, with significant estimates (5\%-level) throughout. For the MRAR$(2)$ model, we again do not only observe worse~AIC and~BIC, but also a non-significant $\alpha_2$-estimate. This shows that an AR$(1)$-structure indeed seems to be sufficient to describe the Swedish population data, which agrees with an analogous conclusion by \citet{kachour09}. The competing RAR$(1)$ model, however, does not only have slightly worse AIC and BIC values, it also does not allow to approximate the s.\,e.\ due to a singular Hessian matrix. So altogether, the MRAR$(1)$ model is clearly preferable among the candidate models. Finally, like in Section~\ref{Financial Time Series}, its adequacy is analyzed based on the standardized Pearson residuals. Their sample acf does not show significant values, their mean $\approx 0.000$ is very close to zero, and their variance $\approx 1.056$ is close to one, confirming that the Swedish population rates data are well represented by the MRAR$(1)$ model.

\begin{figure}[h]
\centering\small
(a)\hspace{-3ex}\includegraphics[viewport=0 45 405 235, clip=, scale=0.65]{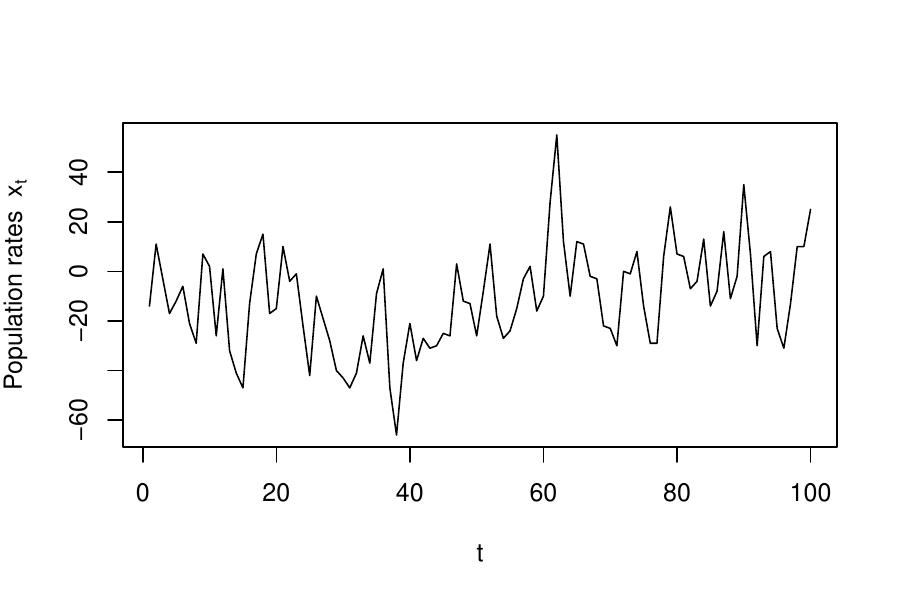}\hspace{-1ex}$t$
\qquad
(b)\hspace{-3ex}\includegraphics[viewport=0 45 190 235, clip=, scale=0.65]{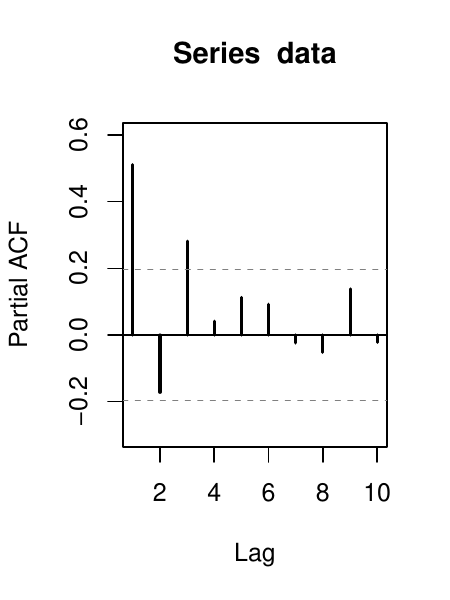}\ $k$
\caption{Temperature index data from Section~\ref{Ecological Time Series}: (a) time series plot and (b) sample pacf against lag~$k$.}
\label{figTemperature}
\end{figure}

\begin{table}[htbp!]
\centering
\caption{Temperature index data from Section~\ref{Ecological Time Series}: ML estimates (approximate s.\,e.\ in parentheses) and information criteria for different candidate models.}
\label{tabTemperature}
$$
\begin{array}{l|@{\qquad}rrrrrr@{\qquad}cc}
\toprule
\text{Skellam-} & \lambda_1 & \lambda_2 & \alpha_1 & \alpha_2 & \alpha_3 & \alpha_4 & \text{AIC} & \text{BIC} \\
\midrule
\text{\iid} & 198.975 & 210.125 &  &  &  &  & 889.2 & 894.4 \\
 & \text{\footnotesize (21.624)} & \text{\footnotesize (21.627)} &  &  &  &  &  &  \\
\midrule
\text{MRAR}(1) & 150.176 & 155.229 & 0.527 &  &  &  & 860.8 & 868.6 \\
 & \text{\footnotesize (29.284)} & \text{\footnotesize (29.272)} & \text{\footnotesize (0.093)} &  &  &  &  &  \\
\midrule
\text{MRAR}(2) & 145.024 & 151.348 & 0.620 & -0.179 &  &  & 858.7 & 869.1 \\
 & \text{\footnotesize (16.616)} & \text{\footnotesize (16.617)} & \text{\footnotesize (0.049)} & \text{\footnotesize (0.050)} &  &  &  &  \\
\midrule
\text{MRAR}(3) & 130.415 & 134.463 & 0.681 & -0.365 & 0.308 &  & 851.7 & 864.8 \\
 & \text{\footnotesize (15.937)} & \text{\footnotesize (15.935)} & \text{\footnotesize (0.096)} & \text{\footnotesize (0.116)} & \text{\footnotesize (0.101)} &  &  &  \\
\midrule
\text{MRAR}(4) & 146.211 & 150.104 & 0.677 & -0.356 & 0.296 & 0.018 & 855.2 & 870.9 \\
 & \text{\footnotesize (22.036)} & \text{\footnotesize (22.033)} & \text{\footnotesize (0.051)} & \text{\footnotesize (0.067)} & \text{\footnotesize (0.127)} & \text{\footnotesize (0.075)} &  &  \\
\midrule
\text{RAR}(3) & 131.585 & 136.049 & 0.647 & -0.340 & 0.647 &  & 852.7 & 865.7 \\
 & \text{\footnotesize (---)} & \text{\footnotesize (---)} & \text{\footnotesize (---)} & \text{\footnotesize (---)} & \text{\footnotesize (---)} &  &  &  \\
\bottomrule
\end{array}
$$
\end{table}

\subsection{Ecological Time Series}
\label{Ecological Time Series}
As our final data example, we consider a time series of annual zonal means (zone ``EQU-24N'') of the land-ocean temperature index (in steps of 0.01 degrees Celsius) for the period 1880--1979 (hence $n=100$). The data are offered by GISS Surface Temperature Analysis (GISTEMP), version 4, of the NASA Goddard Institute for Space Studies. The data have been accessed 2023-11-27 at \url{https://data.giss.nasa.gov/gistemp/} and, thus, constitute an updated version of the data analyzed by \citet[Section~2.4.3]{li24}. The time series is plotted in Figure~\ref{figTemperature}\,(a), and its sample pacf in Figure~\ref{figTemperature}\,(b) now indicates a higher-order autoregressive model structure. As the pacf values at lags~1 and~3 are significant, an AR$(3)$-type model should be appropriate for the data, which agrees with the conclusions in \citet{li24}. In view of these pacf results, we use Skellam-MRAR$(p)$ candidate models up to order $p=4$, and as a further competitor, we also consider the Skellam-RAR$(3)$ model of \citet{kachour14}. Note that the preferred choice by \citet{li24}, a so-called ``INARS$(p)$ model'' fitted by CLS estimation, is not fully parametrized. As the conditional mean of this model agrees with the one of the MRAR$(p)$ model, the CLS estimates of INARS$(p)$ and MRAR$(p)$ are identical.
\medskip

The estimation results (with approximate s.\,e.\ in parentheses) in Table~\ref{tabTemperature} confirm our previous conclusions. AIC and BIC select the MRAR$(3)$ model, and also regarding the (non)significance of estimates, only model orders $p\leq 3$ are justified while the $\alpha_4$-estimate is not significant. Compared to the RAR$(3)$ model, the MRAR$(3)$ model has lower values of AIC and BIC, and in particular, it allows to approximate the s\,e. The adequacy of the MRAR$(3)$ model is confirmed by the corresponding Pearson residuals: their sample acf does not show significant values, their mean $\approx 0.000$ is very close to zero, and their variance $\approx 1.010$ is very close to one. Hence, to sum up, also for $\bbz$-valued time series with higher-order dependence structure, the novel MRARMA class turns out to be a suitable approach for model fitting.

\section{Conclusions and Future Research}
\label{Conclusions}
In this paper, we rediscover the mean-preserving rounding operator and propose a corresponding class of $\bbz$-valued ARMA models.
The existence of stationary solutions and important stochastic properties are derived. The various advantages of the novel model class compared to existing ones are demonstrated. These advantages are further confirmed by several real-world data examples, where the fitted MRARMA models show an appealing performance.
\medskip

We conclude with some suggestions for possible future research.
First, one could consider a semiparametric model specification in analogy to \citet{liu21} to provide more flexibility for model fitting. Second, one could construct INGARCH-type model like in \citet{liu13} to provide another alternative to existing INGARCH models. Finally, it would be interesting to develop multivariate, spatial, or spatio-temporal extensions of the MRARMA class.

\subsubsection*{Acknowledgements}
The authors are grateful to Professor Marcelo Bourguignon (Universidade Federal do Rio Grande do Norte, Brazil) for providing various data sets on $\bbz$-valued time series to them.
Zhu's work is supported by National Natural Science Foundation of China (No. 12271206),  and Science and Technology Research Planning Project of Jilin Provincial Department of Education (No.\ JJKH20231122KJ).

\appendix\small
\numberwithin{equation}{section}
\numberwithin{table}{section}

\section{On an Alternative Model Definition}
\label{On an Alternative Model Definition}
In view of the existing literature on integer time series models, another option for model definition would be
\ba
\label{INARMA_type_Definition}
X_t\ =\ \epsilon_t + \big\langle\alpha_1\cdot X_{t-1}\big\rangle+\ldots+\big\langle\alpha_p\cdot X_{t-p}\big\rangle + \big\langle\beta_1\cdot\epsilon_{t-1}\big\rangle +\ldots + \big\langle\beta_q\cdot\epsilon_{t-q}\big\rangle,
\ea
\ie different rounding operators (executed independently of each other) are used for each part of the model recursion. This type of model definition is inspired by thinning-based ARMA models (``INARMA models''), see \citet{weiss18} for comprehensive information and references on such models. We use the acronym MRARMA\textsuperscript{*} for this kind of modified MRARMA model. Except for low model orders with $p+q\leq 1$, the definitions \eqref{Kachour_type_Definition} and \eqref{INARMA_type_Definition} differ from each other. INARMA models are quite popular for modeling integer time series, so the MRARMA\textsuperscript{*} definition \eqref{INARMA_type_Definition} might look more familiar at a first glance. However, as we shall see in the remainder of this appendix, our original definition \eqref{Kachour_type_Definition} leads to considerably simpler expressions for important stochastic properties.

\begin{figure}[t]
\centering\footnotesize
\includegraphics[viewport=0 45 550 230, clip=, scale=0.7]{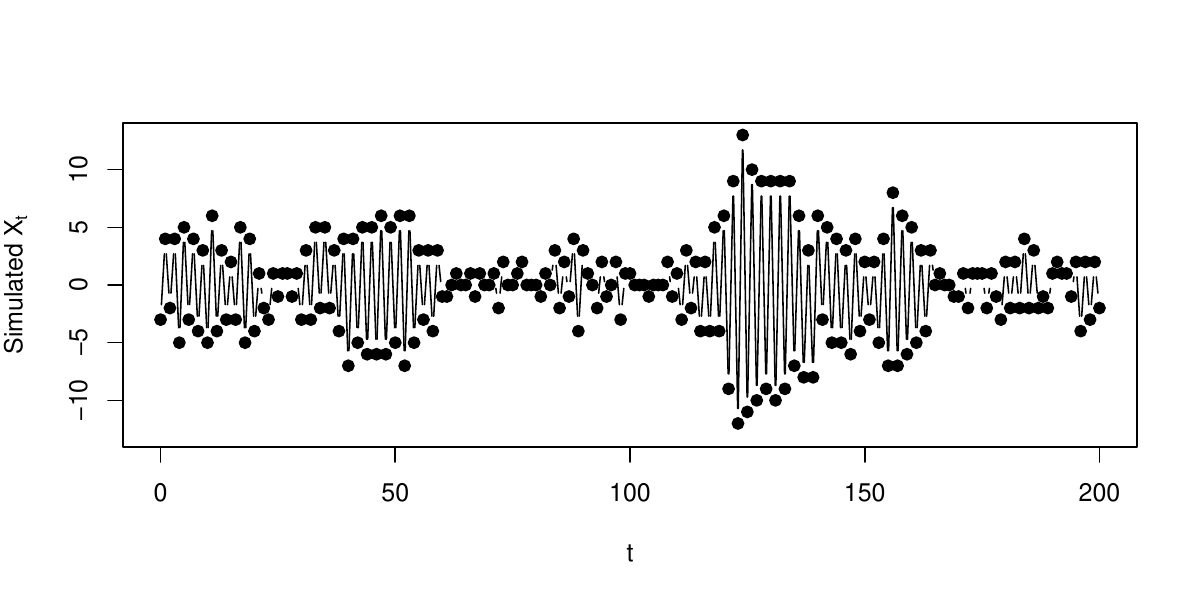}$t$
\caption{Simulated sample path of MRAR\textsuperscript{*}$(2)$-type model according to \eqref{INARMA_type_Definition} with $\alpha_1=-0.6$, $\alpha_2=0.3$, and Skellam innovations with $(\mu_\epsilon,\sigma_\epsilon^2)=(0,2)$.}
\label{figSimRRINAR2}
\end{figure}

\medskip
But let us first start with an important commonness: both model definitions \eqref{Kachour_type_Definition} and \eqref{INARMA_type_Definition} lead to the same linear conditional mean \eqref{CMean}, which implies that they also share the same unconditional mean \eqref{Mean} as well as the same Yule--Walker equations \eqref{Cov} and \eqref{Cov2} concerning the autocovariance structure.
Although \eqref{Kachour_type_Definition} and \eqref{INARMA_type_Definition} have the same (conditional) means, their conditional variances differ. While model \eqref{Kachour_type_Definition} has the simple expression \eqref{CVar_Kachour}, the conditional variance gets more complex for the MRARMA\textsuperscript{*} model \eqref{INARMA_type_Definition}:
\ba
\label{CVar_INARMA}
V[X_t\ |\ X_{t-1},\ldots,\ \epsilon_{t-1},\ldots]\ =\ \displaystyle
\sigma_\epsilon^2 + \sum_{i=1}^p \widetilde{\alpha_i X_{t-i}}\,\Big(1-\widetilde{\alpha_i X_{t-i}}\Big) + \sum_{j=1}^q \widetilde{\beta_j \epsilon_{t-j}}\,\Big(1-\widetilde{\beta_j \epsilon_{t-j}}\Big).
\ea
This additional complexity in conditional variance is one of the reasons why we prefer definition \eqref{Kachour_type_Definition} over \eqref{INARMA_type_Definition}. Note that the heteroscedasticity in \eqref{CVar_INARMA} is more pronounced than in \eqref{CVar_Kachour}, because each summand after~$\sigma_\epsilon^2$ takes a value from $[0,0.25]$, \ie $\sigma_\epsilon^2$ may be increased by up to~$(p+q)\cdot 0.25$. An illustrative simulated sample path is shown in Figure~\ref{figSimRRINAR2}.

\medskip
Next, we turn to the conditional distribution of purely autoregressive MRAR\textsuperscript{*}$(p)$ models according to \eqref{INARMA_type_Definition}.
While \eqref{Kachour_type_Definition} and \eqref{INARMA_type_Definition} agree in the AR$(1)$-case, the transition probabilities get more complex than in \eqref{transprob_Kachour} if $p\geq 2$. In analogy to higher-order INAR models, the conditional distribution is computed by multiple convolutions, which is illustrated for the special case $p=2$:
$$
\begin{array}{rl}
P(X_t=x\ |\ X_{t-1}=x_1,\ldots)
\ =& \sum\limits_{k=-\infty}^\infty P(\epsilon_t=x-k)\,  P\big(\langle \alpha_1 x_1\rangle + \langle \alpha_2 x_2\rangle=k\big)
\\[1ex]
\ =& \sum\limits_{k,l=-\infty}^\infty P(\epsilon_t=x-k)\,  P\big(\langle \alpha_1 x_1\rangle=k-l\big)\,  P\big(\langle \alpha_2 x_2\rangle=l\big).
\end{array}
$$
Considering that $k-l\in (\alpha_1 x_1-1, \alpha_1 x_1+1)$ and $l\in (\alpha_2 x_2-1, \alpha_2 x_2+1)$, the double summation can again be simplified considerably, but it remains more complex than in \eqref{transprob_Kachour}:
\begin{align*}
&P(X_t=x\ |\ X_{t-1}=x_1,\ldots)
\\[1ex]
&\quad =\
\big(1-\widetilde{\alpha_2 x_2}\big)\, \sum\limits_{k=-\infty}^\infty P(\epsilon_t=x-k)\,  P\big(\langle \alpha_1 x_1\rangle=k-\lfloor \alpha_2 x_2\rfloor\big)
\\[1ex]
&\qquad +\ \widetilde{\alpha_2 x_2}\, \sum\limits_{k=-\infty}^\infty P(\epsilon_t=x-k)\,  P\big(\langle \alpha_1 x_1\rangle=k-\lfloor \alpha_2 x_2\rfloor-1\big)
\\[2ex]
&\quad =\
\big(1-\widetilde{\alpha_1 x_1}\big)\big(1-\widetilde{\alpha_2 x_2}\big)\cdot P\big(\epsilon_t=x-\lfloor \alpha_1 x_1\rfloor-\lfloor \alpha_2 x_2\rfloor\big)
\\[1ex]
&\qquad +\ \Big(\widetilde{\alpha_1 x_1}\big(1-\widetilde{\alpha_2 x_2}\big) + \widetilde{\alpha_2 x_2}\big(1-\widetilde{\alpha_1 x_1}\big)\Big)\cdot P\big(\epsilon_t=x-\lfloor \alpha_1 x_1\rfloor-\lfloor \alpha_2 x_2\rfloor-1\big)
\\[1ex]
&\qquad +\ \widetilde{\alpha_1 x_1}\, \widetilde{\alpha_2 x_2}\cdot P\big(\epsilon_t=x-\lfloor \alpha_1 x_1\rfloor-\lfloor \alpha_2 x_2\rfloor-2\big).
\end{align*}

In fact, for general $p\in\bbn$, the conditional distribution is most easily summarized in terms of its conditional pgf,
\ba
\label{condpgf_INARtype}
\e\big[s^{X_t}\ |\ X_{t-1}=x_1,\ldots\big]\ =\ \pgf_\epsilon(s)\cdot \pgf_{\langle \alpha_1 x_1\rangle}(s)\cdots \pgf_{\langle \alpha_p x_p\rangle}(s),
\ea
where the factors $\pgf_{\langle \alpha_i x_i\rangle}(s)$ are computed via \eqref{rounding_operator_pgf}.
Altogether, our original model proposal \eqref{Kachour_type_Definition} is more feasible, which is the reason why the MRARMA\textsuperscript{*} model \eqref{INARMA_type_Definition} is not further discussed in this article.

\section{Proof of Theorem \ref{existence}}
\label{Proofs}
We use arguments similar to those in Theorem~1 of \cite{kachour14}.  Note that we can also establish existence and ergodicity of the process \eqref{Kachour_type_Definition} $(X_t)$ by using
Theorem~3.3 of \citet{menshikov94}. For all $\fy_0\in E^\ast$, we define the empirical measure $\lambda_n$ by
$$
\lambda_n(\cdot)=\tfrac{1}{n}\big[\pi^1(\fy_0,\cdot)+\ldots\pi^n(\fy_0,\cdot)\big],
$$
where $\pi^n$ denotes the $n$-step transition probability of the Markov chain $(\fY_t)$. We shall prove that the sequence of probability distributions $(\lambda_n)$ has a sub-sequence converging to some probability distribution $\lambda$, and this limit will
automatically be a stationary probability distributions of the Markov chain $(\fY_t)$.
Then, the process \eqref{Kachour_type_Definition} can be written in the following form
\begin{equation}\label{matrxi-form}
\fY_t= \mA\,\fY_{t-1}+\feta_t+\fB_t,
\end{equation}
where $\fB_t=\left(\langle Z_{t-1}\rangle-Z_{t-1},0,\ldots,0\right)^\top$. Note that $\|\fB_t\|_1\leq1$. Iterating equation \eqref{matrxi-form}, we have
$$
\textstyle
\fY_t = \mA^n\,\fY_{t-n}+\sum\limits_{j=0}^{n-1} \mA^j\,\fxi_{t-j},
$$
where $\fxi_t=\feta_t+\fB_t$. Define the function $V$ as $V(\fy)=\|\fy\|_1$. Since $V$ is positive and as $k\geq 1$, we have
$\lim_{\|\fyi\|_1\rightarrow\infty} V(\fy)=\infty$, so $V$ is  a Lyapunov function \citep[see][p.~41]{duflo97}. Then,
$$
\textstyle
(\pi^n V)(\fy)=E_{\fyi}\left[\Big\|\mA^n\, \fY_{t-n}+\sum\limits_{j=0}^{n-1} \mA^j\,\fxi_{t-j}\Big\|_1\right]^k.
$$
Note that $\|\fxi_{t-j}\|_1 \leq 2|\epsilon_{t-j}|+1$, $0\leq j\leq n-1$. Let $b=\big(\e[2|\epsilon_{t-j}|+1]^k\big)^{1/k}$, then it follows that $b<\infty$ as $E|\epsilon_{t}|^k<\infty$.
We denote by $\vvvert\cdot\vvvert_1$ the matrical norm
associated to $\|\cdot\|_1$. Thus, from Minkowski's inequality, we obtain
$$
[(\pi^n V)(\fy)]^{1/k}\ \leq\ \vvvert \mA^n\vvvert_1\ \|\fy\|_1+S_n(\mA)\,b,
$$
where $S_n(\mA)=\vvvert \mA^{n-1}\vvvert_1+\ldots+\vvvert \mA\vvvert_1+1$. From Gelfand's formula, we know that for any matrix norm $\vvvert\cdot\vvvert$, we have
$\vvvert \mA^n\vvvert^{1/n} \rightarrow \rho(\mA)$ as $n\rightarrow\infty$. Recall that $\rho(\mA)\leq\vvvert \mA\vvvert$ for any matrix norm $\vvvert\cdot\vvvert$. As $\rho(\mA)<1$,
there exists an $n_0$ such that for any $n\geq n_0$, we have $\vvvert \mA^{n}\vvvert_1<1$. On the other hand, for $n\geq n_0$, we have $\vvvert \mA^n\vvvert_1 \leq (\rho(\mA)+\varepsilon)^n$ with $\varepsilon$ being arbitrarily small and $\rho(\mA)+\varepsilon<1$. Then, $S_n(\mA)$ is convergent using the property of geometric series.

For $n=n_0$, we have $\vvvert \mA^{n_0}\vvvert_1=\alpha^\ast<1$. Then,
$$
\big[(\pi^{n_0} V)(\fy)\big]^{1/k}\ \leq\ \alpha^\ast\|\fy\|_1+\beta^\ast\ =\ \|\fy\|_1\left(\alpha^\ast+\frac{\beta^\ast}{\|\fy\|_1}\right),
$$
where $\beta^\ast=b\sum_{j=0}^\infty\vvvert \mA^{j}\vvvert_1$ and $0<\beta^\ast<\infty$. It follows that
$$
\frac{\big[(\pi^{n_0} V)(\fy)\big]^{1/k}}{V(\fy)}\ \leq\ \left(\alpha^\ast+\frac{\beta^\ast}{\|\fy\|_1}\right)^k.
$$
Then, $(\fY_t)$ satisfies the Lyapunov criterion with $V(\fy)$ as Lyapunov function, see Proposition~2.1.6 in \cite{duflo97}. Moreover, the law of the \iid\ noise $\epsilon_t$ charges all points of $E^\ast$, so $(\fY_t)$ is irreducible. Thus, $(\fY_t)$ is positive recurrent with a unique invariant probability measure, denoted by $\lambda$, with $\lambda(V)<\infty$. On the other hand, part (2) of the theorem follows from the classical ergodic theorem for Markov chains.

\end{document}